\documentclass[twocolumn,showpacs,english,preprintnumbers,amsmath,amssymb]{revtex4}
\usepackage[T1]{fontenc}
\usepackage[latin1]{inputenc}
\usepackage{amsmath}
\usepackage{graphicx}
\usepackage{amssymb}
\usepackage{esint}

\makeatletter
\@ifundefined{textcolor}{}
{%
 \definecolor{BLACK}{gray}{0}
 \definecolor{WHITE}{gray}{1}
 \definecolor{RED}{rgb}{1,0,0}
 \definecolor{GREEN}{rgb}{0,1,0}
 \definecolor{BLUE}{rgb}{0,0,1}
 \definecolor{CYAN}{cmyk}{1,0,0,0}
 \definecolor{MAGENTA}{cmyk}{0,1,0,0}
 \definecolor{YELLOW}{cmyk}{0,0,1,0}
 }

\usepackage{epsf}

\makeatother

\makeatother

\makeatother

\makeatother

\usepackage{babel}

\makeatother

\usepackage{babel}

\begin{document}

\title{Concomitant Modulated Superfluidity In Polarized Fermi Gases }

\author{L. O. Baksmaty$^{1}$, Hong Lu$^{1}$, C. J. Bolech$^{2,1}$ and Han
Pu$^{1}$ }

\affiliation{$^{1}$Department of Physics and Astronomy and Rice Quantum Institute, Rice University, Houston, TX 77005, USA}

\affiliation{$^{2}$Department of Physics, University of Cincinnati, Cincinnati, OH 45221, USA}
\begin{abstract}
Recent groundbreaking experiments studying the effects of spin polarization
on pairing in unitary Fermi gases encountered mutual qualitative and quantitative
discrepancies which seem to be a function of the confining geometry.
Using novel numerical algorithms we study
the solution space for a 3-dimensional fully self-consistent formulation of realistic systems with up to $10^{5}$ atoms.
A study of the three types of solutions obtained demonstrates a tendency
towards metastability as the confining geometry is elongated. One
of these solutions, which is consistent with Rice experiments at high trap aspect
ratio, supports a state strikingly similar to the long sought Fulde-Ferrel-Larkin-Ovchinnikov state. Our study helps to resolve the long-standing controversy concerning the discrepancies between the findings from two different experimental groups and highlights the versatility of actual-size numerical calculations for investigating inhomogeneous fermionic superfluids.
\end{abstract}

\date{\today}
\pacs{03.75.Ss, 71.10.Ca, 37.10.Gh}

\maketitle
Superfluidity in a system of fermionic particles occurs when bosonic
degrees of freedom emerge and condense via pairing of fermions. Understanding
the strength of this pairing mechanism is closely tied to the search
for high temperature superconductors. A central
issue that has animated this quest is: what happens to the pairs when
the participating species have mismatched Fermi surfaces? Such a scenario
occurs, for example, in the presence of a polarizing field, or when the pairing
species have unequal numbers or masses. The issue being that, when the mismatch of the Fermi
surface is large enough, a competition between a normal polarized
state and the superfluid state would ensue \cite{clogston,chandra},
potentially giving rise to yet unknown or poorly understood exotic
superfluid states \cite{casalbuoni}. Among the interesting theoretical
proposals for $s$-wave pairing is the Fulde-Ferrel-Larkin-Ovchinnikov (FFLO) state,
a collective term for an inhomogeneous superfluid referring either
to a Fulde-Ferrell state (FF) \cite{FF_original} which supports a
supercurrent, or a Larkin-Ovchinnikov state (LO) \cite{LO_original}
with a spatially modulated order parameter. Other proposals include
breached pairing and $p$-wave symbiotic superfluids \cite{casalbuoni}.
Primarily due to lack of sufficient experimental evidence, the issue
remains largely unresolved even though it is central to many forms
of matter such as superconductors, neutron stars and color superfluids in the quark-gluon plasma \cite{casalbuoni}. Ultra-cold samples of two component degenerate
Fermi gases \cite{Jin,mit_science,rice_science,rice_stoof_prl,mit_nature}
have re-energized the debate because of their exquisite controllability.

In this paper, we focus on apparently contradictory results on spin-imbalanced unitary Fermi gases from recent experiments between two leading groups \cite{mit_science,mit_nature,rice_science,rice_stoof_prl}. 
In both experiments it was observed that, consistent with earlier
predictions \cite{chandra,clogston}, the trapped superfliud responds to polarization
by phase separating into an inner core with negligible polarization
surrounded by a polarized outer shell. However, the Rice experiments
\cite{rice_science,rice_stoof_prl} performed in cigar-shaped traps
with total particle numbers $N\sim10^{5}$ observed a significant
and unexpected deformation of the central superfluid core, indicating a clear violation of the local density approximation (LDA). In addition,
these results also suggest a much higher superfluid to normal (Chandrasekhar-Clogston)
transition than the MIT experiments \cite{mit_science,mit_nature} 
in which no deformations were observed. The excellent quantitative
agreement with theory \cite{Lobo_QMC,drummond_3d} for
the MIT experiments conducted at much lower trap aspect ratio and with
higher particle numbers $N\sim10^{6}$, hints that there might be
unexpected physics at work in the Rice experiment. In addition, the
concurrence of experiments performed in Paris \cite{Salomon} with
the MIT experiments also suggest a crucial role of the trapping geometry.
This impasse has inspired speculation about the possible role of exotic
phases such as the FFLO state in the observed discrepancies, and stirred much discussion and debate over the past few years by the cold atom community.

The apparent contradiction between the Rice and MIT experiments reflects theoretical difficulties within
trapped geometries: since the effective chemical potential $(\mu)$
varies in space,
several phases may co-exist within a trapped sample. Consequently, despite
excitement and considerable effort, the theoretical complexity inherent
within the problem has ensured that most treatments have, with few
exceptions \cite{machida,Ueda_tezuka,sensarma,adilet,pei}, invoked the LDA \cite{Mueller_lda,duan} which is not general enough
to capture states such as the FFLO. Although an intriguing LDA treatment which
phenomenologically includes a surface energy correction has been able
to account for the shape of the distortions \cite{Mueller_surfacet},
further studies reveal that this model is not consistent with a microscopic
calculation of the surface tension \cite{mueller_bdr}. On the other
hand, recent studies employing variational techniques in isotropic
geometries \cite{yoshida-yip,bulgac_fflo} have shown that the region
of stability for the FFLO state is much larger than originally predicted
\cite{Sheehy}. Until now, a fully self-consistent treatment
in anistropic geometries with realistic particle numbers, has been
well out of reach despite its relevance here and in a wide variety
of other physical systems. To surmount this problem, we developed
scalable numerical techniques which take full advantage of today's
high-performance computing facilities running parallel codes over
thousands of CPUs.

We consider a gas of spin-polarized fermionic atoms confined to a
harmonic trap defined in cylindrical coordinates $(r,\phi,z)$ by
$V(r,\phi,z)=\frac{m}{2}(\omega_{\bot}^{2}r^{2}+\omega_{z}^{2}z^{2})$
with axial and radial frequencies denoted by ($\omega_{z},\omega_{\bot})$.
Consistent with Ref. \cite{rice_science,rice_stoof_prl} we work at the unitarity limit where the $s$-wave scattering length between the two spin species
($a_{s}$) diverges and within a cigar-shaped trap with aspect
ratio defined by $\alpha=\omega_{\bot}/\omega_{z}$. This system of
$N=N_{\uparrow}+N_{\downarrow}$ atoms is described by a Hamiltonian
$\hat{H}=\int d\vec{r}\:(\hat{H}_{0}+\hat{H}_{I})$ with non-interacting $(\hat{H}_{0})$
and interaction $(\hat{H}_{I})$ energy densities given by:
\begin{eqnarray}
\hat{H}_{0}(\vec{r}) & = & \sum_{\sigma=\uparrow,\downarrow}\psi_{\sigma}^{\dagger}\left[{-\frac{\hbar^{2}}{2m}\nabla^{2}+V\left(r,z\right)-\mu_{\sigma}}\right]\psi_{\sigma}\nonumber \\
\hat{H}_{I}(\vec{r}) & = & -U \, \psi_{\uparrow}^{\dagger}(\vec{r})\psi_{\downarrow}^{\dagger}(\vec{r})\psi_{\downarrow}(\vec{r})\psi_{\uparrow}(\vec{r})\,,\label{eq:basic_hamiltonian}\end{eqnarray}
 where $\psi_{\sigma}(\vec{r})$ and $\psi^\dag_{\sigma}(\vec{r})$ represents the fermionic field operators,
$m$ the mass and $\mu_{\sigma}$ the chemical potential of atomic
species with spin $\sigma$. Henceforth, we work in trap units for
which: $m=\omega_{z}=\hbar=1$. The bare coupling constant $U$
is re-normalized through a relationship with $a_{s}$ by: $1/U=-1/4\pi a_{s}+(1/V_{l})\sum1/2\epsilon_{k}$
\cite{drummond_3d,Castin_bcs_theory}, with $\epsilon_{k}=k^{2}/2$
and $V_{l}$ represents the system volume. $\hat{H}$ is diagonalized
through the Bogoliubov-de Gennes (BdG) formulation. In particular, our formulation is
identical to that in Ref.~\cite{drummond_3d}. The superfluid gap (order
parameter) is defined by: $\Delta(\vec{r})=U\langle\psi_{\uparrow}(\vec{r})\psi_{\downarrow}(\vec{r})\rangle$
and the spin densities are given by: $\rho_{\sigma}(\vec{r})=\langle\psi_{\sigma}^{\dagger}\left(\vec{r}\right)\psi_{\sigma}\left(\vec{r}\right)\rangle$.
We find it clarifying to express our results in terms of the Fermi
energy $E_{\textit{F}}=(3N)^{1/3}\alpha^{2/3}$, 
and the Thomas-Fermi radius along the $z$-axis $Z_{\textit{F}}=\sqrt{2E_{\textit{F}}}$
for a single species ideal Fermi gas of $N/2$ particles in a trap
with identical parameters. In addition, following a convention that
$N_{\uparrow}>N_{\downarrow}$, we define $k_{\textit{F}}^{\uparrow\downarrow}=\sqrt{2\mu_{\uparrow\downarrow}}$
and the FFLO wave number by $q_{\,0}=k_{F}^{\uparrow}-k_{F}^{\downarrow}$.

We solve the BdG equations \cite{drummond_3d} using a piece-wise
linear finite-element basis which yield sparse matrices amenable to
efficient parallelization and work in a canonical formalism which
fixes $N$ and the total polarization $P=(N_{\uparrow}-N_{\downarrow})/N$.
It has been recently shown that, in the particular circumstances of
the Rice experiment \cite{rice_science,rice_stoof_prl}, evaporative cooling
shortens the major axis ($z$-axis) of what should be an ellipsoidal
partially polarized region, where the condensate forms \cite{Parish_transport}.
Starting from an initial ansatz for the gap $(\Delta_{I})$ imitating this circumstance,
the BdG equations are iteratively solved to self-consistency using
a modified Broyden's method \cite{Johnson}. Our calculations reveal
that: 
(1)
For large particle numbers ($N\gtrsim 10^4$), we always find a solution similar in structure
to the LDA solution which has the lowest free energy. However starting
from an axially shortened initial ansatz for the gap, this solution is not
accessed by the iterative procedure. 
(2) The most likely solution which is consistent with the Rice experiment is a metastable state that supports a partially polarized superfluid phase strikingly similar to the FFLO phase. This state becomes increasingly robust as trapping geometry becomes more elongated.
(3) Even within a trapped environment, the nodes of the order parameter in the FFLO-like phase are radially aligned which, with low enough noise, provides a measurable, incontrovertible signal within the density profiles.

Superfluidity, a phenomenon of quantum rigidity, acquires its name
from a scenario in which due to energy barriers, a condensate gets
indefinitely trapped within a current-carrying metastable state. The
portent for the experiments under discussion is that the observed
state could be a long-lived metastable state. Thus, we take the approach
of exploring the solution space using ans\"{a}tze constructed with reference
to \cite{Parish_transport} and the phase diagram on the BCS side
of the Feshbach resonance \cite{parish_phase_diagram,Sheehy}. Specifically,
we use the LDA solution for the gap $(\Delta_{{\rm LDA}})$ as a base
to construct an initial ansatz $\Delta_{I}$ which is axially partitioned
into different regions:
\begin{equation}
\Delta_{I}(r,z)=\begin{cases}
\Delta_{{\rm LDA}} & ;\;|z|<z_{c}\\
\Delta_{{\rm LDA}}\cos[q(z-z_{c})] \,e^{-(z-z_{c})^{2}/\lambda^{2}} & ;\;|z|>z_{c}.\end{cases} \nonumber \end{equation}
$\Delta_{I}$ allows us to explore various distorted states. In its
most general form, one encounters the unpolarized BCS,
FFLO and normal phases as one traverses along the axial
direction from the trap center to the edge. The initial size of the
FFLO region in the ansatz is determined by $\lambda$. When $\lambda$ is too small
to accomodate a single wavelength of the gap oscillation, i.e., $0<\lambda<2\pi/q$,
we start without an FFLO phase and $z_{c}$ represents the axial coordinate
of superfluid to normal (S/N) transition. Conversely, an FFLO phase
is initially present in the ansatz when $\lambda>2\pi/q$. In this case $z_{c}$
represents the superfluid to FFLO (S/FFLO) transition. Henceforth
we refer to these initial conditions as $\Delta_{I}^{{\rm P-N}}$
and $\Delta_{I}^{{\rm P-SF}}$, respectively, which reflects our nomenclature
for the eventual solutions as well, i.e., we name the entire solution
according to the character of the partially polarized region: We have
a partially polarized superfluid solution (P-SF) when there is an
FFLO-like phase present. When the partially polarized region is completely
normal, we refer to the entire solution as a P-N solution. For clarity
we single out the LDA-like solution which is obtained when $\Delta_{I}=\Delta_{\rm LDA}$,
as the SF solution. In both the P-N and P-SF solutions, the central unpolarized BCS superfluid core is shortened along the $z$-axis in comparison to the LDA-like SF solution. As we shall see, this shortened BCS core is manifested in the LDA-violating distortion of the density profile of the the minority spin component.

A broad feature of our results, which directly informs on the question
of metastability, is the observation of a barrier between the shortened
states (either P-N or P-SF) and the SF solution. For small atom numbers, this barrier is
absent, the converged solution is unique, independent of the initial
ansatz we take, and we see a dramatic departure from the LDA prediction due to significant finite-size effect.
However with increasing $N$, the axial S/N or S/FFLO transition point
is pinned near its initial value $z_{c}$ and we obtain different solutions
by starting from different initial ans\"{a}tze. Starting from $\Delta_{I}^{{\rm P-SF}}$
or $\Delta_{I}^{{\rm P-N}}$ we always converge to a shortened state
in a manner which is {\em only} sensitive to our choice of $q$.
In other words, we do see a transition between the P-SF and P-N states
which is very sensitive to $q$ and largely insensitive to $\lambda$;
both of which are set in the initial condition $\Delta_{I}$. It works
as follows. When $q$ is less than a critical value $q_{c}$,
the oscillations in the ansatz $\Delta_{I}$ are amplified and the
solution flows to a P-SF state no matter the size of $\lambda$. Conversely
when $q>q_{c}$, the oscillations are damped and $\Delta_{I}$ always
converges to a P-N state. A similar resonance behavior has also recently
been observed in studies of the S/N boundary while tuning $a_{s}$
across the BEC-BCS crossover \cite{mueller_bdr}, in which case calculations
were performed without the radial confinement. It is possible that
this phenomenon might be exploited to engineer the realization of
the P-SF state. 

\begin{figure}
\includegraphics[width=3.4in]{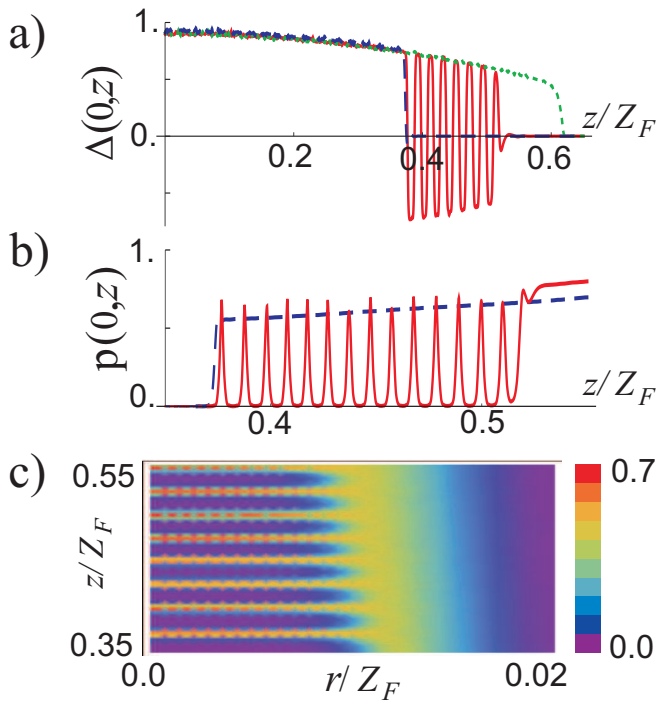}
\caption{\label{fig:z-Profiles} (Color online) (a) Axial profiles
of the gap (in units of $E_F$) showing the P-N (blue dashed line), P-SF (red solid line) and 
SF (green dotted line) states. The LDA solution (not shown)
almost completely overlaps with the SF result. The free energies per
particle are: $0.67(0)E_{\textit{F}}$, $0.65(8)E_{\textit{F}}$, $0.65(5)E_{\textit{F}}$
and $0.64(4)E_{\textit{F}}$ for the P-N, P-SF, SF and LDA states,
respectively. (b) Local polarization $p(\vec{r})$ within the partially
polarized region of the P-SF(red solid line) and P-N(blue dashed line) solutions. (c) An
$r$-$z$ plot of the normalized density difference $\delta\rho=(\rho_{\uparrow}-\rho_{\downarrow})/\rho_{F}$
of the partially polarized region of the P-SF state $(\rho_{\textit{F}}=\sqrt{(2E_{F})^{3}}/6\pi^{2})$. All the results shown in this paper are obtained at a small temperature $T=0.02 E_F/k_B$, and with $N=50000$, $\alpha=50$,
and $P=0.3$.}
\end{figure}

We ascribe the consistent convergence to a shortened state as due
to the emergence of energy barriers separating the P-SF and P-N states
from the SF state with increasing $N$ or $\alpha$ in tandem with
$E_{F}$. In Fig.~\ref{fig:z-Profiles}(a) we illustrate the dramatic
differences in the superfluid gap for the various solutions encountered.
Apart from the emerging energy barriers, another important result
with regard to metastability is the decrease in the relative energetic
separation of all the states, P-SF, P-N and SF, as $\alpha$ is increased.
Taken together, these observations suggest that the relaxation of
the physical system from any of the shortened states to the SF state, which is the lowest in energy, becomes {\em
less} favorable as $\alpha$ is increased, a deduction which is borne
out by the discrepancies of the Rice and MIT experiments.

For a given value of $z_{c}$, the energy of the P-SF solution is consistently lower
than the P-N solution. Furthermore, recent results suggest that the inclusion
of fluctuations, neglected in mean-field formulations, should make
the P-SF state even more stable \cite{bulgac_fflo}. Thus, we expect that
if the system converges to a shortened state, it will choose the P-SF
state. A natural question to ask is: how will the FFLO phase manifests
itself?

\begin{figure}[h]
\includegraphics[width=3.2in]{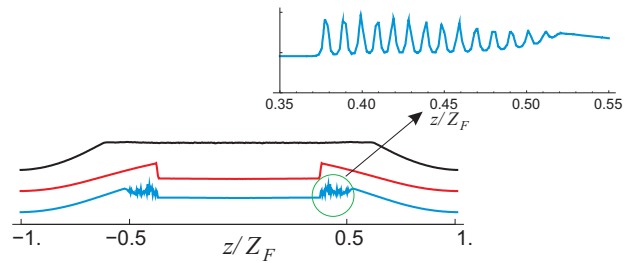}
\caption{\label{fig:dblstack}(color online) Plots showing the doubly integrated  axial spin density 
$\delta\rho_{1d}(z)$ for, from top down, the SF, P-N and P-SF states
shown in Fig.~\ref{fig:z-Profiles}(a). 
}
\end{figure}

In Fig.~\ref{fig:z-Profiles}(b) we contrast
the appearance of local polarization $p(\vec{r})=(\rho_{\uparrow}-\rho_{\downarrow})/(\rho_{\uparrow}+\rho_{\downarrow})$
in the partially polarized regions of the P-SF and P-N states. We
note that in \cite{bulgac_fflo}, the strong oscillations displayed
in $p(\vec{r})$ were observed to survive the effects of fluctuations.
One pleasant surprise of our results was the radial alignment of the
nodes of the FFLO phase shown in Fig.~\ref{fig:z-Profiles}(c). A fact
which is not {\em a priori} obvious, and very promising for the
prospects of detection within the 3D system under discussion here,
because it implies that the FFLO phase could yield a measurable signal
in the density profile. Auspiciously,
it also suggests that when an array of 1D tubes, such as are being
used in current experiments \cite{yean_liao}, are coupled to yield
a quasi-3D confinement, the FFLO nodes at each tube are likely to
align to yield a measurable signal. To make sure that the radial alignment of the nodes is not a numerical artefact, we have used initial ans\"{a}tze where the nodes are intentionally misaligned along the radial direction. Our code always converges to states with the nodes aligned. A comparison of the plots in Fig.~\ref{fig:dblstack}
confirms that the presence of an FFLO phase would indeed provide a
smoking gun signal in doubly integrated axial spin density $\delta\rho_{1d}=\int \int dxdy (\rho_{\uparrow}-\rho_{\downarrow})$. In the close-up we observe
that the signal of the FFLO region is not as strong as that in Fig.~\ref{fig:z-Profiles}(c)
because of contributions from the fully polarized shell encasing it.
Quantitatively, it indicates that a lower bound of the signal to noise
ratio $\approx 6.5$ is required to observe at least half of the
FFLO phase. 

\begin{figure*}
\includegraphics[width=5.5in]{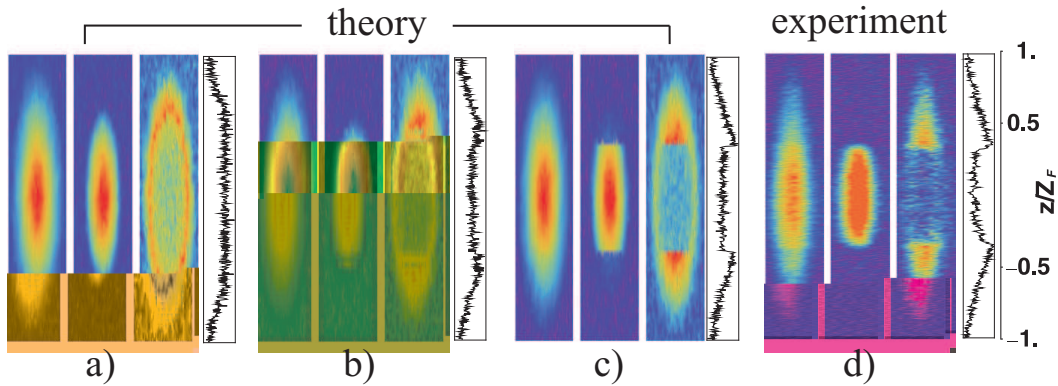}
\caption{\label{fig:bcs_column} (color online) Each display shows the column densities (rescaled to have aspect ratio 5 for clarity) $\int \rho_{\uparrow}dx$, $\int \rho_{\downarrow}dx$, $\int (\rho_{\uparrow}-\rho_{\downarrow})dx$
and the axial spin density $\delta \rho_{1d}$,
respectively. The states represented are (a) the SF, (b) P-SF and
(c) P-N states illustrated in Fig.~\ref{fig:z-Profiles}(a). In (d)
we plot the Rice experimental results for $N\approx 260000$, $P\approx 0.35$,
$\alpha=45.23$ and $T<0.05 E_F/k_B$.}
\end{figure*}

A casual comparison of all column density profiles in Fig.~\ref{fig:bcs_column}
rules out the observation in Rice experiment of the SF state, which is consistent with the LDA and, within the BdG formulation, has the lowest free energy.
However, due to the noise on the experimental data, it is not clear
which of the shortened states (P-SF or P-N) has been observed. To
produce the noise with similar characteristics as the experiment,
we added white noise with standard deviation which is a similar fraction
of the average value of the column density $\int_{-\infty}^{\infty}\rho_{\uparrow\:}dx$
in the plotted window. Theoretically, since it has the lower energy and since the transition between the FFLO phase and the normal phase is continuous,
one expects that, between the two shortened states, the P-SF solution will
be favored. 

In conclusion, we have repeatedly solved the BdG equations in a cigar-shaped
trap using initial conditions which imitate the condensate nucleation
process \cite{Parish_transport}. The iterative solution chooses between
two stationary points, which are not necessarily the global free energy minimum,
but each of which features density profiles strikingly similar to experimental observations at Rice.
The solution which possesses the lower energy of these two contains
an FFLO-like phase which leaves an accessible signal in $\delta\rho_{1d}$.
Coupled with recent results which suggest the unexpected stability
of the FFLO in 3D \cite{bulgac_fflo}, our observations raise the interesting
question of whether the FFLO state has already been realized in the Rice experiment. Since
the Hartree interactions are excluded from the BdG formulation of unitary gases \cite{drummond_3d},
we do not address the position of the Clogston limit. Nevertheless
we note in passing that the P-SF solution has the capacity to absorb
polarizations and, if undetected, could conceal the existence of a
partially polarized region. Finally we remark that our work is important for another reason,
as far as we know ours are by far the largest calculations of their
kind and herald the arrival of an important tool for investigations
of finite fermionic systems such as occur in atomic traps or in nuclear
physics; where predictions of ideal models such as the FFLO proposal
could be significantly modified by confinement and finite-size effects. 

{\em Note added} After our work was completed, the Rice experimental group verified the suggestion made in Ref.~\cite{Parish_transport} that the LDA-violating deformations observed in their experiment are a result of depolarization of the superfluid core by evaporation occurring mainly at the axial center of the trap \cite{randyc}. They found that these deformed states are very stable, in agreement with our calculations. The metastability of these states suggests the possiblity to directly engineer an FFLO state in an elongated trap.

We thank U. Landman for computing resources at the Center For Computational
Material Science and R. Barnett for insight on the numerical procedures.
Part of this work was performed at NERSC, Navy DSRC and the ARSC.
We also thank E. Mueller, R. Hulet, D. Huse, S. Pollack, H. Hu, M. Forbes and
S. Bhongale for many illuminating discussions. This work was supported by a grant from the ARO with funding from the DARPA OLE program, the Welch foundation (C-1669, C-1681) and NSF.

\end{document}